\documentstyle[aps,epsfig,floats,amssymb]{revtex}

\begin{document}
\frenchspacing
\twocolumn[\hsize\textwidth\columnwidth\hsize\csname
@twocolumnfalse\endcsname
\title{Spinor Gravity}
\author{A. Hebecker$^*$,
C. Wetterich$^{**}$}
\address{
$^{*}$Deutsches Elektronen-Synchrotron, Notkestra\ss e. 85, D-22603 Hamburg,
Germany\\
$^{**}$Institut f{\"u}r Theoretische Physik, Philosophenweg 16, D-69120
Heidelberg, Germany}
\maketitle

\begin{abstract}
A unified description of all interactions could be based on a
higher-dimensional theory involving only spinor fields. The metric arises
as a composite object and the gravitational field equations contain
torsion-corrections as compared to Einstein gravity. Lorentz symmetry in
spinor space is only global, implying new goldstone-boson-like gravitational
particles beyond the graviton. However, the Schwarzschild and Friedman
solutions are unaffected at one loop order. Our generalized gravity seems
compatible with all present observations.
\end{abstract}

\pacs{PACS numbers:
12.10.-g;
04.20.Cv;
11.10.Kk
\hfill DESY-03-090,\hspace{.5cm}HD-THEP-03-31}

]

The physical description of our world is based on fermionic and bosonic
degrees of freedom, corresponding to particles with different statistical
properties. In particular, bosons mediate the interactions between
fermions as well as between themselves. One may consider two alternative
approaches to a fundamental unified theory. Either one treats bosons and
fermions on equal footing, attempting to link them by some symmetry
principle - this is the road of supersymmetry. Alternatively, one considers
the fermions as fundamental and explains the bosons as composite objects of
an even number of fermions. The second approach is commonly followed in the
treatment of strongly correlated electrons in statistical physics and is
familiar from the Nambu-Jona-Lasinio~\cite{NJL} and Gross-Neveu~\cite{gn}
models\footnote{There are also approaches of treating fermions as composite
objects in a bosonic theory and one can imagine a dual description of a
fundamental theory (see, e.g.,~\cite{sky}).}.

In this letter we explore the hypothesis that a fundamental unified
theory of the gravitational, electroweak and strong interactions can
be formulated in terms of only fermionic degrees of freedom. Given that
all bosons arise as composite fields, one immediately faces the fundamental
challenge of describing the metric in terms of fermionic
variables. In particular, one has to realize the symmetry of general
coordinate transformations or diffeomorphisms by a purely fermionic
functional integral {\em without} the use of a fundamental metric degree
of freedom. The metric of space-time can then be associated with the
expectation value of a suitable fermion bilinear. Our formulation uses the
notion of $d$-dimensional coordinates $x^\mu,$ $\mu\!=\!0,\dots,d\!-\!1$, only
in order to describe spinor fields $\psi(x)$ in some local coordinate patch
and for the definition of a derivative $\partial_\mu\psi$. The geometric
and topological properties of space-time arise a posteriori as a
consequence of the dynamics of the system, reflected in the fermion
correlators. This realizes the idea of obtaining geometry from general
statistics \cite{GenS}.

The idea of a purely fermionic fundamental theory has previously been
explored on the basis of proposals for ``pregeometry''~\cite{acmt,ak} or
``metric from matter''~\cite{av}, where a diffeomorphism invariant spinor
theory was formulated and the metric was obtained as a composite object.
However, in~\cite{acmt} and most of the subsequent work on pregeometry, an
(auxiliary) metric or vielbein field had to be introduced already at the
level of the fundamental action. In the spirit of ``induced
gravity''~\cite{sak} (see~\cite{adl} for a review), fermionic loop
corrections then generate the kinetic terms for these fields, transforming
them into propagating degrees of freedom. By contrast, in~\cite{av} a purely
spinorial action was proposed. However, it is of non-polynomial form
(the covariant derivative involves terms $\sim (\psi\bar{\psi})^{-1}$).
Since elements of the inverse matrix $(\psi\bar{\psi})^{-1}$ do not exist 
within the Grassmann algebra, it is not clear how a functional integral 
can be defined. A later supersymmetric proposal~\cite{luk} involves again 
fundamental bosonic fields. 

The proposal coming closest to our ideas about `spinor gravity' is contained
in Akama's pregeometry paper~\cite{ak}, where a purely spinorial, polynomial
action with diffeomorphism invariance is given. However, after introducing
bosonic auxiliary fields a singular limit is taken to realize local
Lorentz symmetry. This limit is inconsistent with polynomiality of the
original action. As will become clear below, our proposal is free of such
problems because we do not insist on local Lorentz invariance at a
fundamental level. Exploring the consequences of a spinor theory with only
global Lorentz symmetry will lead us to a generalized gravity theory which
is, in principle, distinguishable from Einstein gravity but appears to
be consistent with all present experimental bounds.

Once an effective gravity theory is found, the extension to gauge
interactions and scalar fields (needed for spontaneous symmetry breaking)
is straightforward: one may go higher-dimensional~\cite{KK}. Indeed, it is
well known that non-abelian gauge interactions arise in the compactification
of pure higher-dimensional gravity theories~\cite{dew}. In contrast to
supersymmetric
unified models, the number of dimensions is not restricted here to $d\leq
11$. This opens the door for realizing the gauge symmetries of the standard
model as isometries of ``internal space'', with light fermions protected by
chirality. Rather realistic models of quarks and leptons with electroweak
and strong interactions have already been proposed in the context of
18-dimensional gravity coupled to a Majorana Weyl spinor \cite{CWQL}.

Spinors are Grassmann variables and our basic tool is the Grassmann
functional integral for the partition function
\begin{equation}\label{1}
Z[J]=\int{\cal D}\psi \exp\Big\{-(S+S_J)\Big\}.
\end{equation}
The dynamics is described by the ``classical action'' $S[\psi]$ which has to
be a polynomial functional of the spinor fields.
As a technical tool we have added to the action a source term $S_J$. It is
linear in suitable sources $J$ and facilitates the computation of
expectation values of spinor bilinears. We will set $J=0$ in the end to
obtain expectation values in the vacuum. Our first task is to find a
polynomial action $S$ which is invariant under diffeomorphisms, so that no
particular choice of coordinates is singled out. For this purpose we may use
as a basic building block the ``real'' local fermion bilinears
\begin{equation}\label{2}
\tilde{E}^m_\mu(x)=\frac{i}{2m}\Big\{\bar{\psi}(x)\gamma^m\partial_\mu
\psi(x)-\partial_\mu\bar{\psi}(x)\gamma^m\psi(x)\Big\}.
\end{equation}
Here $\psi(x)$ belongs to an irreducible spinor representation of the
global $d$-dimensional ``Lorentz group'' SO(1,$d\!-\!1$) and $\gamma^m$
are the associated Dirac matrices. The arbitrary mass scale $m$ has only
been introduced to make $\tilde{E}^m_\mu$ dimensionless. Under general
coordinate transformations $\psi$ transforms as a scalar and therefore
$\tilde{E}^m_\mu$ as a vector. Interpreting $\tilde{E}^m_\mu$ as a $d\times
d$ matrix and defining $\tilde{E}=\det(\tilde{E}^m_\mu)$, we propose the
invariant action\footnote{
In
the conventions of Eq.~(\ref{2}) one has $\alpha\sim m^d$. There is,
however, no need to introduce $m$. For dimensionless spinor fields
$\psi(x)$ and omitting the factor $m^{-1}$ in Eq.~(\ref{2}), the
coefficient $\alpha$ becomes dimensionless. It can be scaled to arbitrary
values by a rescaling of $\psi$. We employ here a ``Euclidean notation''
avoiding factors of $i$ in the definition of the functional integral. For a
Minkowski signature a factor $i$ can be absorbed in $\alpha$.
}
\begin{equation}\label{3}
S=\alpha\int d^dx\tilde{E}=\frac{\alpha}{d!}\int d^dx\epsilon^{\mu_1
\dots\mu_d}\epsilon_{m_1\dots m_d}\tilde{E}^{m_1}_{\mu_1}\dots
\tilde{E}^{m_d}_{\mu_d}.
\end{equation}
Clearly, no metric is needed in order to realize diffeomorphism invariance
in a fundamental fermionic theory~\cite{ak}. Since we demand that the action
is polynomial in $\psi$, no inverse of $\tilde{E}^m_\mu$ can appear and all
lower world indices from derivatives $\partial_\mu$ have to be contracted
with the totally antisymmetric $d$-dimensional $\epsilon$-tensor. This
also ensures the cancellation of the Jacobian of reparameterizations.
The action of Eq.~(\ref{3}) is invariant under global but not under local
Lorentz rotations.

More general forms of invariant actions can be found in an accompanying
paper~\cite{LP}. As explained above, they have to contain precisely $d$
derivatives by reparameterization invariance. The number of spinor fields
contained in each invariant may
vary, but it is bounded above due to the finite number of derivatives
and because a sufficiently high power of a spinor field vanishes by
anticommutativity. Thus, only a finite (though potentially large) number
of independent invariants exists. 

The basic composite bosonic field is the ``global vielbein'' $E^m_\mu$
associated with the expectation value of the fermion bilinear
$\tilde{E}^m_\mu$
\begin{equation}\label{4}
E^m_\mu(x)=\langle\tilde{E}^m_\mu(x)\rangle =
\frac{\delta\ln Z}{\delta J^\mu_m(x)}.
\end{equation}
Here we have specified the source term $S_J=-\int d^dx$\linebreak
$J^\mu_m(x)\tilde{E}^m_\mu(x)$. Using $\eta_{mn}=\,\,$diag$(-1,1\dots 1)$,
the metric can be constructed in the standard way
\begin{equation}\label{5}
g_{\mu\nu}(x)=E^m_\mu(x)E_{\nu m}(x)=E^m_\mu(x)E^n_\nu(x)\eta_{mn}.
\end{equation}
The inverse metric $g^{\mu\nu}$ exists whenever $E=\det(E^m_\mu)$\linebreak
$\neq0$ and can be used to ``raise indices''. As required, the metric
transforms as a symmetric tensor field. We will argue that the metric obeys
field equations similar to the Einstein equations. It can be associated with a
massless, composite graviton (see also \cite{comp} for some of the early
as well as more recent related ideas). However, we expect the resulting
effective gravity theory to be a generalization of Einstein gravity. Since
the action of Eq.~(\ref{3}) is invariant under {\em global} but not under
{\em local} Lorentz rotations, the vielbein describes additional massless
degrees of freedom not associated with the metric. (This distinguishes our
approach from earlier attempts to realize local Lorentz
symmetry~\cite{acmt,ak,av}.) The resulting ``generalized
gravity''~\cite{CWGG} will lead to a specific version of torsion. In
suitable dimensions and for a suitable ``ground state'' $E^m_\mu$, our
fundamental spinor theory will describe both massless spinors and massless
gravitational fields.

In order to get a first glance at the dynamics of the theory, we employ the
method of partial bosonization \cite{HS}. Without a perturbative expansion
in a small parameter we do not expect quantitatively accurate results, but
the structural elements deriving from the effective degrees of freedom and
symmetries will become visible. Up to an irrelevant normalization constant,
we can write the partition function as an equivalent functional integral
over fermions $\psi$ and boson fields $\chi$ (cf.~\cite{gn}),
\begin{eqnarray}\label{6}
Z[J]=\int{\cal D}\psi{\cal D}\chi^n_\nu \exp
\Big\{-S_B+\int d^dx J^\mu_m\chi^m_\mu\Big\}\,,\nonumber\\
S_B[\psi,\chi]=\alpha\int d^dx\Big[\det(\tilde{E}^m_\mu)
-\det(\tilde{E}^m_\mu-\chi^m_\mu)\Big].
\end{eqnarray}
The equivalent purely fermionic theory can be recovered by performing the
Gaussian functional integral for the bosonic fields $\chi^n_\nu$. It differs
from the theory defined by Eqs. (\ref{1}) and (\ref{3}) only in the
connected correlation functions of at least order $d$ in
$\tilde{E}_\mu^m$~\cite{LP}. With $E^m_\mu=\langle\chi^m_\mu\rangle$,
the effective action $\Gamma[E]$ is defined as
\begin{eqnarray}\label{7}
\Gamma[E]&=&-W+\int d^dxJ^\mu_m(x)E^m_\mu(x),\\
\frac{\delta W[J]}{\delta J_m^\mu}&=&E_\mu^m\,,\qquad W[J]=\ln Z[J]\,.
\label{8}
\end{eqnarray}
By construction, $\Gamma[E]$ is invariant under diffeomorphisms and global
Lorentz rotations if the functional measure is free of gravitational
anomalies (which could, in principle, be present for
$d=2\,$mod$\,4$~\cite{AW}). The quantum field equations following from the
variation of $\Gamma[E]$ are
\begin{equation}\label{8AA}
\frac{\delta\Gamma}{\delta E^m_\mu}=J^\mu_m\,,
\end{equation}
where $J^\mu_m=0$ in the vacuum. Contributions from matter and radiation
can be associated to an energy momentum tensor $T^{\mu\nu}=E^{-1}E^{m\mu}
J^\nu_m$. The latter accounts for the incoherent fluctuations of $\psi$ and
of bosonic composite fields as well as for possible expectation values
of bosonic composite fields other than $E^m_\mu$.

At one loop order in the fermionic fluctuations, $\Gamma[E]$ is given by
$(\tilde{\alpha}=(-1)^{d+1}\alpha)$
\begin{equation}\label{9}
\Gamma[E]=\tilde{\alpha}\int d^dxE-\frac{1}{2}\,\mbox{Tr}\,\ln(E{\cal D})\,,
\end{equation}
where $E{\cal D}$ is the second functional derivative of $S_B$ with respect
to $\psi$. The generalized Dirac operator ${\cal D}$ involves the inverse
vielbein $E^\mu_m$ and a ``covariant derivative'' $\hat{D}_\mu$,
\begin{eqnarray}\label{10}
{\cal D}&=&\gamma^\mu\hat{D}_\mu=E^\mu_m\gamma^m\hat{D}_\mu~,  \\
\hat{D}_\mu&=&\partial_\mu+\frac{1}{2E}E^m_\mu\partial_\nu(EE^\nu_m).
\label{11}
\end{eqnarray}
The evaluation of the one loop term can be performed by standard techniques
(see, e.g.,~\cite{gus} and references therein). We expand in the number of
derivatives and find (see below) for the term with two derivatives
\begin{eqnarray}\label{12}
\Gamma_{(2)}&=&\frac{\mu}{2}\int d^dxE\Big\{-R\nonumber\\
&&+\tau[D^\mu E^\nu_m D_\mu E^m_\nu-2D^\mu
E^\nu_mD_\nu E^m_\mu]\Big\}\,,
\end{eqnarray}
where $\tau=3$ and
\begin{equation}\label{12aa}
D_\mu E^m_\nu=\partial_\mu E^m_\nu-\Gamma_{\mu\nu}^{\ \ \ \lambda}
E^m_\lambda.
\end{equation}
Here $\Gamma_{\mu\nu}^{\ \ \ \lambda}$ and $R$ are the usual Christoffel
symbols and curvature scalar associated with the
metric $g_{\mu\nu}$. Thus, the present rough approximation suggests that
our spinor model leads to effective gravitational interactions.

Deviations from Einstein gravity result from the term $\sim\tau$. If
$E_\mu^m$ were a {\em local} vielbein of conventional general relativity,
this term would be forbidden by {\em local} Lorentz symmetry since
$D_\mu$ contains no spin connection.
In order to interpret the additional invariant, it is useful to study the
linear expansion around flat space
\begin{equation}\label{13}
E^m_\mu=\delta^m_\mu+\frac{1}{2}(h_{\mu\nu}+a_{\mu\nu})\eta^{\nu m}\,,
\end{equation}
where $h_{\mu\nu}$ and $a_{\mu\nu}$ are the symmetric and antisymmetric
part. The former accounts for the metric fluctuations $(g_{\mu\nu}=
\eta_{\mu\nu}+h_{\mu\nu})$ and can be decomposed in the usual manner,
\begin{eqnarray}\label{14}
h_{\mu\nu}&=&b_{\mu\nu}+\frac{1}{(d-1)}
\left(\eta_{\mu\nu}-\frac{\partial_\mu\partial_\nu}{\partial^2}\right)
\sigma\nonumber\\&&+\frac{\partial_\mu\partial_\nu}{\partial^2}f+
\partial_\mu v_\nu+\partial_\nu v_\mu\,,
\end{eqnarray}
with $b_{\mu\nu}\eta^{\mu\nu}=0$, $\partial_\mu b^{\mu\nu}=0$, and
$\partial_\mu v^\mu=0$ ($v^\mu$ and $f$ being gauge degrees of
freedom). The new degrees of freedom are associated with the antisymmetric
part $a_{\mu\nu}$,
\begin{equation}\label{15}
a_{\mu\nu}=c_{\mu\nu}+\partial_\mu(v_\nu+w_\nu)-\partial_\nu
(v_\mu+w_\mu),
\end{equation}
where $\partial_\mu c^{\mu\nu}=0$ and $\partial_\mu w^\mu=0$. Expanding the
effective action of Eq.~(\ref{12}) to quadratic order yields
\begin{eqnarray}\label{16}
\Gamma_{(2)}&=&\frac{\mu}{8}\int d^dx\left\{\partial^\mu b^{\nu\rho}
\partial_\mu b_{\nu\rho}-
\frac{d-2}{d-1}\partial^\mu\sigma\partial_\mu\sigma\right.\nonumber\\
&&\left.+\tau\partial^\mu c^{\nu\rho}\partial_\mu c_{\nu\rho}\right\}.
\end{eqnarray}
The terms involving $b_{\mu\nu}$ and $\sigma$ are the same as in Einstein
gravity. While $v^\mu$ and $f$ are absent because of diffeomorphism
invariance, the absence of $w^\mu$ may indicate a hidden nonlinear
symmetry. The new massless particles described by $c_{\mu\nu}$ can be
interpreted as the Goldstone bosons of the spontaneous breaking of the
global Lorentz symmetry. Indeed, a ground state $E^m_\mu=\delta^m_\mu$ breaks
the global SO(1,$d\!-\!1$) symmetry acting only on the indices $m$
completely, while it is invariant under combined global Lorentz
transformations acting simultaneously on the coordinates. This implies
$d(d-1)/2$ Goldstone directions corresponding to $a_{\mu\nu}$. In Einstein
gravity with local Lorentz symmetry $a_{\mu\nu}$ would be a gauge degree of
freedom.

Furthermore, on flat space a spinor mass term is forbidden by global
Lorentz rotations for irreducible spinors in
$d\!=\!2,6,8,9\,$mod$\,8$~\cite{CWMS}. In these dimensions the spectrum
contains both massless fermions and massless gravitons, where the latter
can be interpreted as fermionic bound states. In a higher-dimensional
theory $(d>4)$ this is a necessary~\cite{CWMS} condition for obtaining
a realistic low energy theory with quarks and leptons and the gauge
symmetries of the standard model. Indeed, the most interesting solutions
for the vielbein correspond to geometries where the characteristic size of
$d-4$ dimensions is much smaller than the size of the 3+1 dimensional
universe observable on large length scales. The effective four-dimensional
theory is then given by ``compactification''. Part of the higher-dimensional
symmetries (connected to isometries) appear in the effective theory
as gauge symmetries. The chirality index~\cite{CWCI} counting the number
of chiral fermion generations with respect to these gauge symmetries
depends on the particular geometry of ``internal space''. For gravity
in $d=18$ rather realistic settings for symmetries and charges have
already been discussed long ago~\cite{CWQL} and it would be very
interesting to see how such a construction could emerge from the
generalization of gravity proposed in this letter.

We assume that, after compactification, the gravitational part of the
effective four-dimensional action still has the form of Eq.~(\ref{12}), now
with new effective coefficients $\mu_4,\tau_4$. The general form of the
action is restricted by the symmetries, even if additional invariants are
added to the classical action of Eq.~(\ref{3}). In this context we note that
the two terms multiplying $\tau$ in Eq.~(\ref{12}) could, from the point
of view of global Lorentz and diffeomorphism invariance, have coefficients
with a ratio other than $-2$. In other words, there are really three
independent invariants only two of which are actually generated by the
loop calculation. If present, the third invariant would lead to more drastic
deviations from
Einstein gravity~\cite{LP} than the invariant $\sim \tau$ generated in
the one loop approximation. In this letter we assume that some deeper
reason (symmetry?) protects the structure of the one loop expression
Eq.~(\ref{12}) at higher-loop order. As a working hypothesis we take the
coefficient of the other allowed invariant to be zero (or very small).
As a test of the viability of a generalized gravity theory with only global
Lorentz symmetry, we therefore compare the field equations obtained from the
effection action Eq.~(\ref{12}) in $d=4$ with observations. In this context
we also assume that the cosmological constant in the effective 4-dimensional
theory (almost) vanishes. Postponing a more detailed investigation of this
interesting issue, we note that the classical cosmological term in
Eq.~(\ref{9}) can be chosen to cancel the one-loop contribution, which
amounts to the usual fine-tuning. This cancellation also depends on the
internal compact space and extends to a more general form of the
fundamental spinor action which could involve other invariants beyond
Eq.~(\ref{3}).

For weak gravity the linear field equations follow from the variation of
$\Gamma_{(2)}$ in Eq.~(\ref{16}) in presence of an energy momentum tensor
$T_{\mu\nu}$. With energy density
$T_{00}=\rho$ and Newtonian potential $\phi=-\frac{1}{2} h_{00}$, one finds
for static solutions $(\Delta=\partial^i\partial_i)$
\begin{equation}
\label{17}\Delta\phi=\frac{\rho}{2\mu}.
\end{equation}
This associates $\mu$ with the reduced Planck mass
$\bar{M}_p^2=M^2_p/8\pi=(8\pi G_N)^{-1}=\mu$.

The new field $c_{\mu\nu}$ only couples to the antisymmetric part of the
energy momentum tensor, $(T_A)_{\mu\nu}$, defined by the variation of the
action with respect to the antisymmetric part of $E_\mu^m$. As will become
clear below, for spinors this corresponds to a coupling to the spin.
Point-like particles without internal structure (including massless
particles) generate a symmetric energy momentum tensor. Therefore
$(T_A)_{\mu\nu}$ can only be related to the internal structure -- in our case
it involves the spin vector. The ``spin contribution'' to the gravitational
interactions mediated by the exchange of $c_{\nu\rho}$ does not contain a
rotation invariant part since $c_{0i}$ and $\epsilon_{ijk}c_{jk}$ transform
as vectors. Only for macroscopic objects with a nonzero macroscopic spin
vector the modifications of gravity would become observable.

It has been checked \cite{LP} that even beyond the linear approximation the
modifications of gravity $\sim\tau$ neither affect the Schwarzschild
solution nor the Friedmann cosmological solution. We conclude that our
proposal of generalized gravity is compatible with all present tests of
general relativity. Nevertheless, it is conceivable that the presence
of new long range fields beyond the graviton  may lead to new interesting
cosmological solutions, possibly accounting for quintessence \cite{Q}.

To make the relation of spinor gravity to conventional general relativity
more explicit, it is useful to introduce nonlinear fields with local
Lorentz invariance. We write the global vielbein as $E_\mu^m=
e_\mu^n H_n{}^m$, where $e_\mu^n$ is a conventional ``local vielbein'' with
local Lorentz index. The SO$(1,d-1)$ matrix $H_n{}^m$ has one local and one
global index. The new local Lorentz transformations present in the
nonlinear formulation correspond to a reparameterization of the
decomposition of $E_\mu^m$ in $e_\mu^n$ and $H_n{}^m$, leaving $E_\mu^m$
invariant. The nonlinear field $H_n^{\ m}$ characterizes the
new degrees of freedom that are present in addition to those of conventional
gravity. We define here the covariant derivative $D_\mu$ by the
requirement of a covariantly constant local vielbein $e_\mu^m$. This
implies that the covariant derivative of $E_\nu^m$ is given by
$D_\mu E_\nu^m=e_\nu^nD_\mu H_n{}^m$. Here the action
of $D_\mu$ on the first index of $H$ makes use of the spin connection
defined by $e_\mu^m$, while the second, global index is inert. Given that
$E=e=\,\,$det$(e_\mu^m)$, it is now obvious that Eq.~(\ref{12}) describes
conventional gravity covariantly coupled to the non-linear field $H$.

Geometrically, the new degrees of freedom can be characterized by a further
connection $\tilde{\Gamma}$, which is present in addition to the Riemannian
connection $\Gamma$,
\begin{equation}\label{18AA}
\tilde{\Gamma}_{\mu\nu}{^\lambda}=(\partial_\mu E^m_\nu)E^\lambda_m.
\end{equation}
This connection is defined by demanding that $E_\mu^m$
be covariantly constant. The existence of a covariantly constant basis of
vector fields implies that our new connection is curvature-free. It has,
however, non-vanishing torsion
\begin{equation}\label{18a}
T_{\mu\nu\rho}=(\partial_\mu E_\nu^m-\partial_\nu E_\mu^m)E_{\rho m}.
\end{equation}
Thus, in the bosonic formulation, we are dealing with general relativity
with torsion, where the torsion tensor is restricted by the requirement of
a curvature-free Riemann-Cartan connection. More precisely, given a metric
and a torsion tensor such that the Riemann-Cartan connection has zero
curvature, we can always define a global vielbein $E_\mu^m$ by parallel
transporting a frame defined at an arbitrary but fixed point of the manifold.
One can easily check that $E_\mu^m$ will then automatically satisfy
Eq.~(\ref{18a}). Note that such a theory is very different from general
relativity with a generic (unconstrained) torsion tensor.

In the above geometrical formulation, the one-loop calculation leading to
Eq.~(\ref{12}) is particularly easy to understand. The leading fermionic
term in the lagrangian of Eq.~(\ref{6}) reads
\begin{equation}
{\cal L}_\psi\sim\bar{\psi}E^\mu_m\gamma^m\partial_\mu\psi+\mbox{h.c.}
\end{equation}
Using the local Lorentz invariance introduced above, we choose $e_\mu^m=
E_\mu^m$. In this local frame, the coefficients of the spin connection
going with our Riemann-Cartan connection vanish,
$\tilde{\omega}_\mu{}^{ab}=0$. Thus, we have
\begin{equation}
{\cal L}_\psi\sim\bar{\psi}e^\mu_m\gamma^m\tilde{D}_\mu\psi+\mbox{h.c.}\,,
\label{lpsi}
\end{equation}
and the problem is reduced to evaluating the effective action induced by
integrating out a fermion in a Riemann-Cartan background. In fact,
we can introduce a new fermionic variable $\psi'$ defined to coincide
with $\psi$ in the frame where $e_\mu^m=E_\mu^m$ and to transform like
a usual general-relativity spinor under local Lorentz transformations of
$e_\mu^m$. (This corresponds to the field redefinition $\psi=S(H^{-1})\psi'$,
where $S$ denotes the spinor representation.) Now we have a formulation of
partially bosonized spinor gravity with full local Lorentz invariance: The
variables are $\psi'$, $e_\mu^m$ and $T_{\mu\nu\rho}$ (constrained by the
requirement of a curvature-free Riemann-Cartan connection). The quadratic
part of the fermionic action is given by Eq.~(\ref{lpsi}) with $\psi\to
\psi'$, which has local Lorentz-invariance.

The fermionic one-loop contribution to the effective action for $e_\mu^m$
and $T_{\mu\nu\rho}$ is calculated from the square of the Dirac operator
defined by Eq.~(\ref{lpsi}). More specifically, the leading divergences
are characterized by the DeWitt coefficients~\cite{dew} of the relevant
operator (see, e.g.,~\cite{haw}). While the highest divergence comes with
the cosmological constant, the next contribution contains the leading terms
with vielbein derivatives and torsion. Given that only the totally
antisymmetric part of the torsion couples to fermions, it has to be of the
form
\begin{equation}\label{tf}
\Gamma_{(2)}=\frac{\mu}{2}\int d^dx e\left\{-R+\tau' T_{[\mu\nu\rho]}
T^{[\mu\nu\rho]}\right\}\,,
\end{equation}
The value $\tau'\!\equiv\! 3\tau/4\!=\!9/4$ is fixed, e.g., by the
calculation of the relevant DeWitt coefficient in~\cite{gus} (see
also~\cite{yaj}). Although this formally agrees with the result
of~\cite{ak}, the physical meaning is very different since our torsion
variable is constrained. Thus, the second term in Eq.~(\ref{tf}) is the
kinetic term for our new light degrees of freedom rather than a torsion
mass term. In the linear approximation one finds
$T_{[\mu\nu\rho]}\sim\partial_{[\mu}c_{\nu\rho]}$. The source for
$c_{\nu\rho}$ arises from the coupling $\sim c_{\nu\rho}\partial_\mu(
\bar{\psi}\gamma^{[\mu}\gamma^\nu\gamma^{\rho]}\psi)\sim c_{\nu\rho}
\epsilon^{\nu\rho\mu\sigma}\partial_\mu S_\sigma$, which follows from
Eq.~(\ref{lpsi}) with $\psi$ replaced by $\psi'$. Because of the
superposition of the microscopic contributions from individual fermions,
a macroscopic source for $c_{\nu\rho}$ involves the macroscopic total spin
vector $S_\sigma$.
For known objects the total spin is small and the mass contribution to
gravity dominates by far, rendering the macroscopic effects of the exchange
of $c_{\nu\rho}$ unobservable. However, it is easy to check that the
particles described by $c_{\nu\rho}$ are produced in electron-positron
annihilation or similar processes and their $t$-channel exchange affects
electron-positron scattering. Of course, due to the gravitational coupling
strength the cross section is tiny and not observable, just as for the
graviton.

The effective action of Eq.~(\ref{tf}) depends on $H$ only via the term
quadratic in the totally antisymmetric part of the torsion
$T_{[\mu\nu\rho]}$. Any solution $e^m_\mu(x)$ of Einstein gravity is
therefore  also a solution of the generalized field equations provided
$T_{[\mu\nu\rho]}=0$. In particular, the solution $E^m_\mu=e^m_\mu$ exists
whenever the vielbein is diagonal in a suitable coordinate system and
Lorentz frame. This holds for the Schwarzschild- and
Robertson-Walker-metrics, explaining the result of the explicit
computation of~\cite{LP}. Similarly, $T_{[\mu\nu\rho]}$ vanishes in every
rotation and parity invariant situation since no totally antisymmetric
invariant tensor exists ($\epsilon_{ijk}$ being odd under parity). This
concludes our argument that the generalized gravity of Eq.~(\ref{12}) cannot
be distinguished from Einstein gravity by present observations. It may be a
surprise that a principle as fundamental as local Lorentz symmetry is so
little tested in practice. Global Lorentz symmetry seems to be a perfectly
viable alternative, despite the presence of new massless fields in the
gravitational sector.

Our ambitious proposal for a unified theory of all interactions stands, of
course, only at the beginning of its development. Nevertheless, we believe
that it offers a possible interesting alternative to superstring theories
for a well-defined theory of quantum gravity. For this it is crucial that,
unless affected by anomalies, spinor gravity is renormalizable: As
explained above, the symmetries allow only a finite number of polynomial
invariants. Thus, if the quantum theory respects the symmetries, all
divergences can be absorbed in a finite number of tree-level coefficients
(which is one of the definitions of a renormalizable field theory).
Furthermore, even though global Lorentz and reparameterization symmetry
allow for a large number of independent invariants, we expect that
additional symmetry requirements can drastically reduce the number of
fundamental parameters. One still
needs to specify a regularized functional measure that preserves the
symmetries. If this can be achieved, spinor gravity is well defined by
an explicit functional integral.  The expectation values, including the
metric, become, in principle, calculable.

\end{document}